\documentclass[journal]{vgtc}                

\usepackage{mathptmx}
\usepackage{graphicx}
\usepackage{times}
\usepackage{epstopdf}
\usepackage[utf8]{inputenc}

\usepackage[bookmarks,backref=true,linkcolor=black]{hyperref} 
\hypersetup{
  pdfauthor = {},
  pdftitle = {},
  pdfsubject = {},
  pdfkeywords = {},
  colorlinks=true,
  linkcolor= black,
  citecolor= black,
  pageanchor=true,
  urlcolor = black,
  plainpages = false,
  linktocpage
}

\onlineid{0}

\vgtccategory{Research}

\vgtcinsertpkg

\title{MetroViz: Visual Analysis of Public Transportation Data}

\author{Fan Du, Joshua Brulé, Peter Enns, Varun Manjunatha and Yoav Segev}
\authorfooter{
\item F. Du, J. Brulé, V. Manjunatha and Y Segev are graduate students at the University of Maryland, College Park. E-mail: $\left\{fan|jbrule|varunm|segev\right\}$@cs.umd.edu.
\item P. Enns is affiliated with the University of Maryland Institute for Advanced Computer Studies
. E-mail: peter@umiacs.umd.edu.
}

\abstract{
Understanding the quality and usage of public transportation resources is important for schedule optimization and resource allocation. Ridership and adherence are the two main dimensions for evaluating the quality of service. Using Automatic Vehicle Location (AVL), Automatic Passenger Count (APC), and Global Positioning System (GPS) data, ridership data and adherence data of public transportation can be collected. In this paper, we discuss the development of a visualization tool for exploring public transportation data. 
We introduce ``map view'' and ``route view'' to help users locate stops in the context of geography and route information. To visualize ridership and adherence information over several years, we introduce ``calendar view'' - a miniaturized calendar that provides an overview of data where users can interactively select specific days to explore individual trips and stops (``trip subview'' and ``stop subview'').
MetroViz was evaluated via a series of usability tests that included researchers from the Center for Advanced Transportation Technology (CATT) and students from the University of Maryland - College Park in which test participants used the tool to explore three years of bus transit data from Blacksburg, Virginia.
}

\keywords{Public transportation, visual analysis, spatio-temporal data}

  \teaser{
  \centering
  \includegraphics[width=16cm]{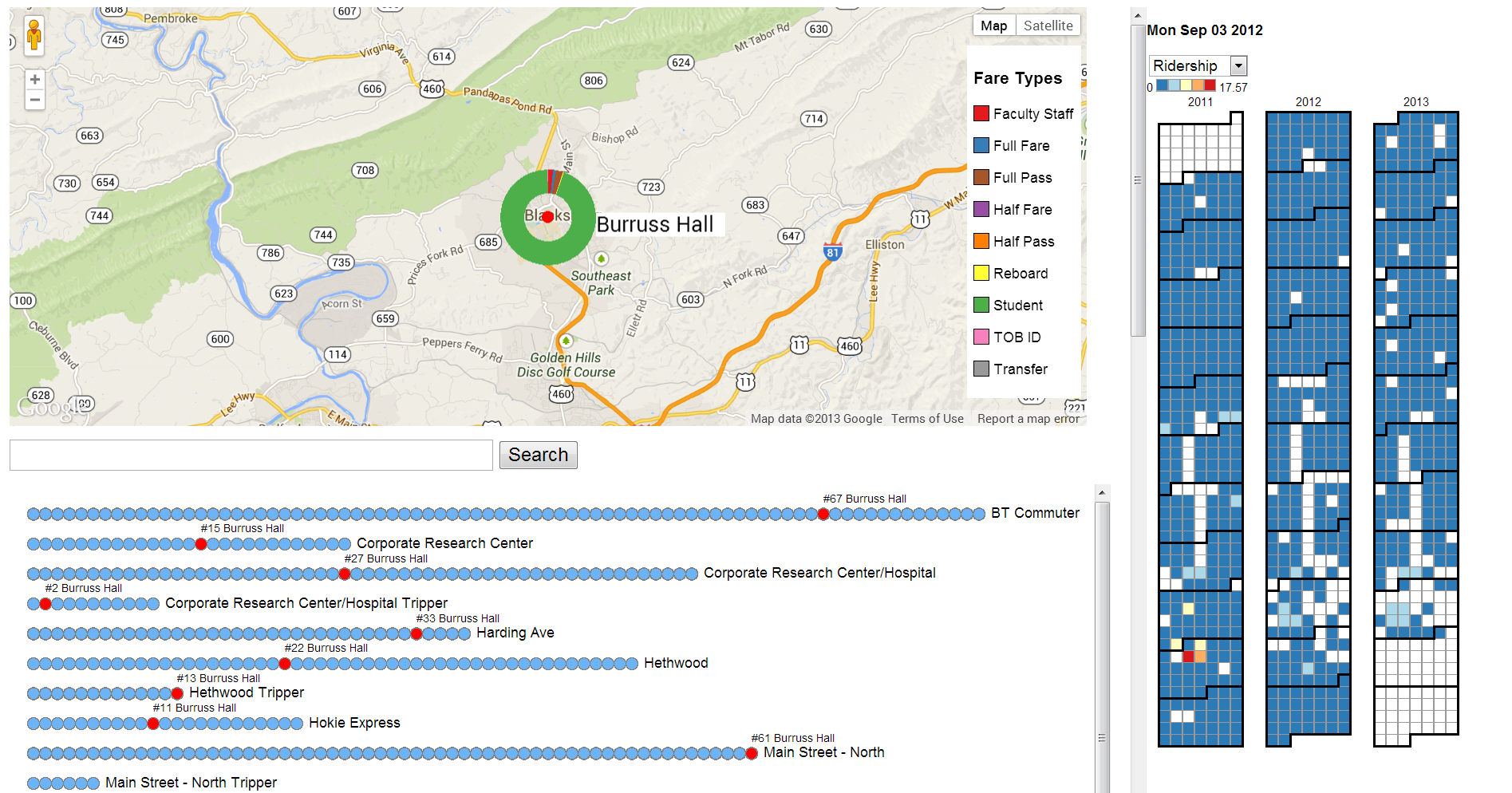}
  \caption{MetroViz begins with a broad overview of any public transportation dataset with its map view and route view.}
  \label{map_view}
  }
\renewcommand{\manuscriptnotetxt}{}

\begin{document}


\firstsection{Introduction}

\maketitle

Public transportation follows scheduled routes and timetables, and allows passengers to board and alight at specific stops or stations. Most research regarding AVL, APC and GPS data focuses on developing algorithms to predict the travel time \cite{Lee:2012:HNF:2424321.2424357,Tiesyte:2008:SPT:1463434.1463452,Predic2007}, to automatically optimize scheduling \cite{4658140,Yu2007}, and to estimate the quality of service \cite{camus2005estimation,hammerle2005use}. To the best of our knowledge, there is no existing visualization tool to aid service providers in interactively exploring \emph{all} of this data with the same tool to help gain insight using the advantages of their domain knowledge.

There are three technical challenges that we believe are critical to visualize the ridership (the number of passengers on board a bus at any given time) and adherence (how closely a bus follows the schedule) data of public transportation over time. The first challenge is to design a general data model that can be applied to different public transportation modes. The second challenge is to design the interface to show the stops/stations, routes, trips, ridership and adherence simultaneously. The third challenge is to determine how to process the large-scale time series data and develop an interactive visual analysis tool.

To address these challenges, we have developed MetroViz, an interactive visual analysis tool that helps users explore public transportation data, and evaluate how well the public transportation system is serving the public. 
We model the public transportation data by separating it into three levels: 1) the stop/station level, 2) the route level and 3) the trip level. As most public transportation modes follow scheduled routes and timetables, and only allow passengers to board and alight at specific stops or stations, our model can be applied to many different modes of public transportation.
Based on the three level model, our visual design consists of three views: the map view, the route view and the calendar view.
We present a map and route view to help users locate stops in the context of geography and route information. To convey ridership and adherence information over years, we present a calendar view inspired by \cite{wicklin2009congestion}. The calendar provides an overview of the data from which users can interactively select specific days to explore the details in the trip and stop subviews of the calendar view.

During usability testing, researchers from the Center for Advanced Transportation Technology (CATT), and students from the University of Maryland - College Park took part in a study in which they used our tool to explore the bus transit data of Blacksburg, Virginia. The participants were able to use our tool to explore bus stops, bus routes, and ridership and adherence data of bus trips.

To the best of our knowledge, our work is the first to help users visually analyze public transportation data, as well as to allow users to interactively explore the large-scale time series ridership and adherence data over time. Our work presents three technical contributions.

\begin{itemize}
  \item \textbf{A general data model} that divides the public transportation data into three levels. This model can be applied to different public transportation modes.
  \item \textbf{An overview-to-detail interface} consists of three views that visualize stops/stations, routes, trips, ridership and adherence data simultaneously.
  \item \textbf{An iterative and progressive visual analysis} that accelerates the query response and enables users to explore the data interactively.
\end{itemize}

Following a review of related work, we describe MetroViz's system design, data processing and visual design of the overview-to-detail interface followed by a description of the usability testing to evaluate the design.

\section{Related Works}

\subsection{Analysis of Public Transportation Data}
Automatic Vehicle Location (AVL), Automatic Passenger Count (APC) and Global Positioning System (GPS) has been used to collect ridership and adherence data - the two main factors for evaluating the quality of service of public transportation.
Rancic et al. \cite{Rancic2008} present a tracking system using AVL data to analyzing city bus transit traffic. 
Chen et al.\cite{5203406} introduce a model to simulate bus operation and passenger demand based on AVL and APC data. 
Camus et al. \cite{camus2005estimation} propose a way to estimation the quality of service based on AVL data. 
Hammerle et al. \cite{hammerle2005use} analyze the AVL and APC data of Chicago Transit Authority to estimate service reliability.
Mai et al. \cite{mai2011visualizing} extend the Marey graph by adding schedule adherence and passenger load information to measure transit performance.
Kimpel et al. \cite{4658140} discuss efforts of using the TriMet APC and AVL data to improve buses scheduling.
Yu and Yang \cite{Yu2007} develop a dynamic holding strategy to optimize the holding strategy.

Public transportation data has also be used to predict the travel time and optimize the route choosing. 
Lee et al. \cite{Lee:2012:HNF:2424321.2424357} introduce a real-time travel time prediction method based on multiple samples of similar historical trajectory. 
Tiesyte and Jensen \cite{Tiesyte:2008:SPT:1463434.1463452} propose a method to predict the future movement of a vehicle based on the identification of the most similar historical trajectory. 
Predic et al. \cite{Predic2007} use real-time AVL data and historical data to predict bus motion and bus arrival time. 
Nguyen et al. \cite{Nguyen2012} treat buses as moving objects, and use temporal maps to represent the movements of buses in spatio-temporal domain to help passengers choose appropriate routes. 
Liu et al. \cite{5958130} propose a bus trip planning system to help passengers choose the most appropriate lines and transfers, based on traffic data.

However, most of the research regarding AVL, APC and GPS data focuses on developing algorithms to predict the travel time, to automatically optimize the bus scheduling, or to estimate the quality of service. To the best of our knowledge, there is no existing visual analysis tool that could aid service providers to interactively explore those data and gain insights using the advantages of their domain knowledge.

\subsection{Spatio-temporal Data Visualization}
Transportation data is a special case of spatio-temporal data. Public transportation modes follow scheduled routes and timetables, and allow passengers to board and alight at specific stops or stations. Much previous research regarding spatio-temporal data visualization focus on trajectory drawing. 
Tominski et al. \cite{Tominski2012} use a hybrid 2D/3D display to show the trajectories and associated attributes in their spatio-temporal context.
Scheepens et al. \cite{6065019} improve density maps to help explore trajectories using multiple density fields.
Some research focuses on graphs to visualize the spatio-temporal data.
Cui et al. \cite{4658140} propose an edge-clustering method to reduce edge crossings and visualize geometry graphs.
Guo \cite{5290710} develops a visualization framework to interactively explore large-scale spatial flows.
Wang and Chi \cite{6065020} introduce a focus+context method to visualize metro map on small displaying area of mobile devices.

However, trajectories cannot effectively show data in both the spatio-temporal context and the ridership-adherence context simultaneously. In our design, we choose to use separated overview-to-detail interface to show stops/stations, routes and trips information on different levels of visualization.

\section{Data Modeling}

\begin{figure}[htb]
 \centering
 \includegraphics[width=6cm]{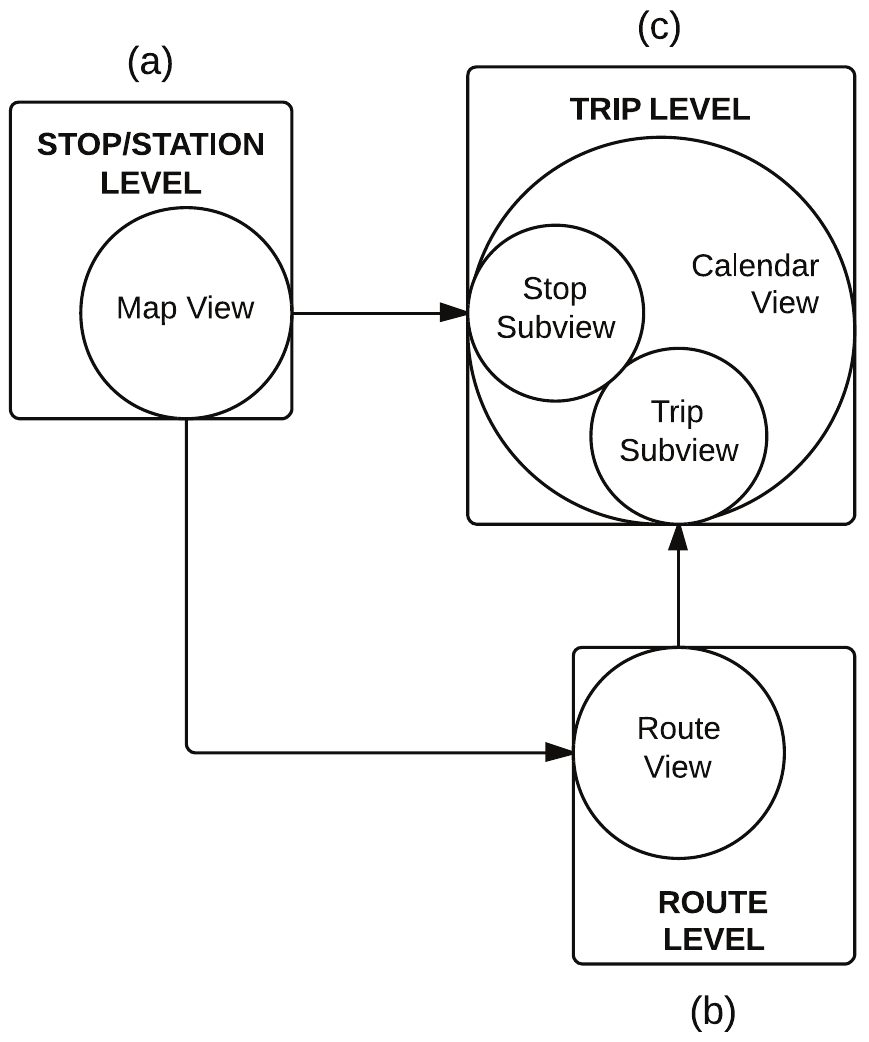}
 \caption{Three level data model for public transportation data. (a) The stop/station level; (b) the route level based on stop/station level; (c) the trip level based on both the stop/station level and the route level.}
 \label{data_model}
\end{figure}

One of the challenges of processing and visualizing public transportation data is how to design a data model that can be applied to different modes of public transportation (e.g., buses, trains, metro, ferries, etc.).
To address this challenge, we modeled the public transportation data by separating it into three levels: the stop/station level (Fig. \ref{data_model}a), the route level (Fig. \ref{data_model}b) and the trip level (Fig. \ref{data_model}c).

Public transportation only allows passengers to board and alight at specific stops/stations. In our project, we model stops/stations as the basic elements and lay them out in the first level. Each stop/station has a specific geographic location, and serves the general public within its covered area.
The route level is based on the stop/station level where each route is scheduled by the service provider and consists of a sequence of stops/stations.
In our model, the trip level shows details of daily operations. A trip is a particular realization of a route, following scheduled timetables. There are two significant aspects for evaluating the quality of service of trips: the ridership (how many passengers a trip serves) and adherence (how well a trip matches its scheduled timetable).

Based on this three level model, our visual design consists of three views: the map view (Fig. \ref{data_model}a), the route view (Fig. \ref{data_model}b) and the calendar view (Fig. \ref{data_model}c). Map view shows public transportation data in stop/station level, highlighting the geographic information for each stop/station, and allows users to explore the data in their geographical context.
Route view shows simplified route information by displaying the stops in their visited order allowing users to explore the stop/station data in the context of the order of the stops.
Finally, calendar view provides daily operation details for both the stop/station level and the route level. Users can explore the ridership and adherence data of a stop/station as well as the complete route. We also provide a stop subview and a trip subview (Fig. \ref{data_model}c) for exploring hourly operation details, as extensions of the calendar view.

\section{Data Processing}

In development of MetroViz, we used three years of bus transit data for the city of Blacksburg, Virginia. The data was collected and provided to us by the Blacksburg Department of Transportation. The data consists of three files - each file includes about three years of data from mid 2010 to mid 2013. Uncompressed, the files totaled over 22 million records and over three gigabytes worth of data.

The first file contains information relating to adherence of buses to their schedules. For every bus route, for every trip on that route, and for every stop on that trip, there is a record of the scheduled departure time from that stop, as well as the actual departure time from that stop, and the difference in seconds between them. A negative difference corresponds to an early arrival, a positive difference corresponds to a late arrival at that stop. The second file consists of data about the number of passengers on the bus at each stop. For every trip and stop, data is available for the number of passengers buying a ticket (collected by the ticketing machines installed on the bus), the number of passengers boarding and the number of passenger alighting the bus (both collected by an automatic counting device, and generally less reliable than the fare count). The third file is a breakdown of the fare counts into categories. Categories includes but not limited to: student fare, faculty/staff, full fare, etc.

The first step we took in processing the data was to write scripts that filter out erroneous records, such as partial or missing records or records that are inconsistent in any way. Inconsistent record include but not limited to: non-matching timestamps between the two files, and invalid route, stop, or trip name. After this, we used another script to load the files into a Sqlite3 database, and used batch SQL queries to align the data of the first file to that of the second file (passengers count, and adherence data). We also loaded the third file (fares breakdown) to the database in order to easily run aggregated statistics. Once the data was ready and quality assured, we exported it to a different set of SQL tables to save memory and improve performance. The latter file is the one used in the visualization process.

To supply the data to the visualization utility, we wrote a simple python based web server using the CherryPy framework that returns JSON formatted data to several types of HTTP requests with a given set of parameters (such as data types, dates, routes, etc.)

\section{System Overview}
MetroViz is organized into three primary views: map view (Fig. \ref{map_view}), route view (Fig. \ref{map_view}) and calendar view (Fig. \ref{calendar_view}). These views are designed to visualize several dimensions of public transportation datasets at different levels of detail, and to allow the user to zoom in on the data they wish to explore and filter out extraneous details.

To begin, the user is presented with a broad overview of the transportation system's organization in space, and routes with two coordinated views (map and route view). In the map view, stops are visualized as blue circles superimposed on a Google map. Mousing over a stop highlights it on both the map and route views and displays the distribution of fare types collected at the stop in a donut glyph around the stop. The route view also displays stops as blue circles. For each route, stops are displayed in linear order that they occur in the actual bus route, followed by the route's name. Mousing over a stop triggers the same response that it would in the map view. Mousing over a route name highlights every stop in the route in both views.

Route view provides two controls to filter data as needed. Clicking on a stop displays all of the data collected at that stop in the calendar view. Similarly, clicking on a route's name displays all of the data collected for the entire route in the calendar view. Calendars in the calendar view are organized in a familiar format. The calendar displays either adherence or ridership for every day where data was collected with the type of data displayed on the calendar controlled via dropdown menu.

\begin{figure*}[htb]
 \centering
 \includegraphics[width=16cm]{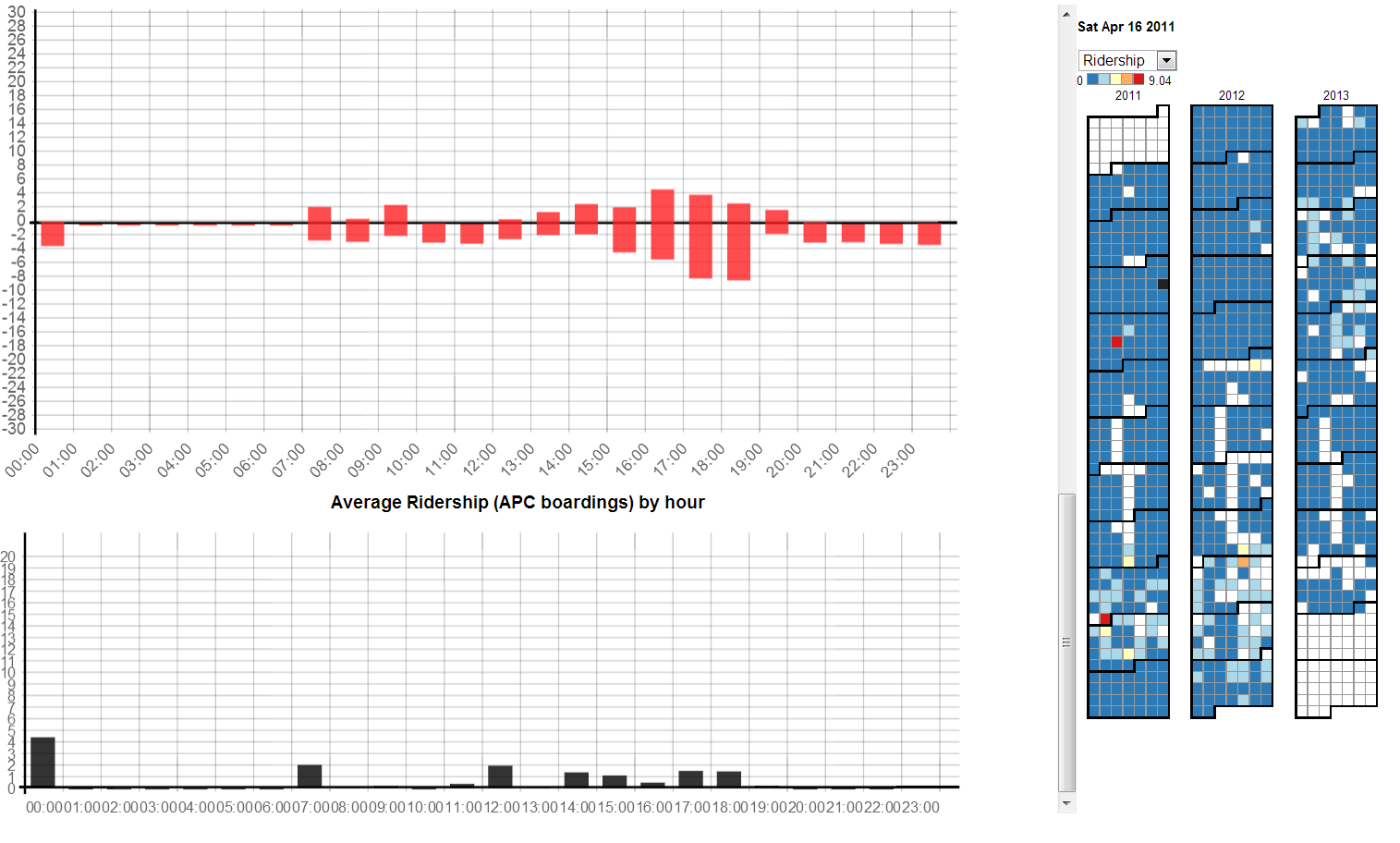}
 \caption{One of MetroViz's detail views (a combination of calendar view and stop subview) can be seen by clicking on a stop and a day on calendar view.}
 \label{calendar_view_big}
\end{figure*}

The user may drill down further from the calendar by clicking on a specific day. If the calendar was generated by clicking on a stop, the stop subview (Fig. \ref{calendar_view_big}) of the calendar view is displayed. Similarly, if the calendar was generated by clicking on a route, the trip subview (Fig. \ref{trip_component}) of the calendar view is displayed. Both of these views provide fine grained details about each stop or route throughout the day.

\section{Visual Design}
\subsection{Map View and Route View}

Map view (Fig. \ref{map_view}) (built on top of Google Maps) is designed to provide an overview the entire Blacksburg bus transit system. Every bus stop in the system appears as a blue dot. Hovering over a stop highlights it in route view (discussed below) and brings up a donut chart summarizing the fare types at that stop. Map view also includes a search feature to search stops by name. Based on requests from the initial usability study, we added drop-down auto-completion to the search box. Entering a stop will pan/zoom the map to center on the appropriate stop. The route view (Fig. \ref{map_view}), placed immediately below the map view, is designed to provide a concise overview of all bus stops in the Blacksburg bus transit system. Each stop is represented by a small blue circle, and a single route is represented by an array of stops. The name of the route is located to the right of the last stop. The Blacksburg bus transit system has 25 active routes which cannot all be displayed without the use of a scroll-bar. The scroll-bar ensures that the user can have a simultaneous view of any route as well the map. The user can interact with route view via the following actions:
\begin{enumerate}
  \item hovering over individual stops (blue circles)
  \item clicking individual stops 
  \item hovering over individual routes (route name text)
  \item clicking individual routes.
\end{enumerate}

The initial design of route view had each stop circle increase in size on mouse-over. During the usability test, several participants requested highlighting instead. In the current version, the stop receives a red highlight and appears as the center of focus in the map view with a red highlight and fare information. The numeric position of the stop in the route, and the name of the stop appear as tooltips. If several routes share a stop, they all receive red highlights with position and name tooltips, in their respective routes. 

When the user clicks a blue circle representing an individual stop, the calendar view for that particular stop (see section 5.2) is produced to the right of the route view.When the user hovers over a route name all stops in that route are highlighted in the route view and map view allow the user to quickly geographically visualize the entire route. When a user clicks the name of a route, the calendar view for that route is generated to the right of the route view.

\subsection{Calendar View}
\begin{figure}[htb]
 \centering
 \includegraphics[width=5cm]{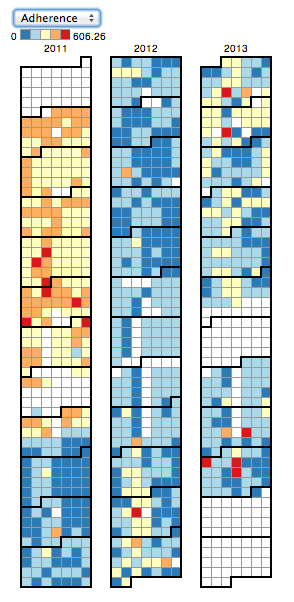}
 \caption{Three years of adherence data displayed on the calendar view. (White squares correspond to missing data.)}
 \label{calendar_view}
\end{figure}

Calendar view (Fig. \ref{calendar_view}) is a series of miniaturized monthly calendars stacked vertically. A single year in  calendar view is displayed as a grid with seven columns and enough rows to display 12 months where each cell represents a day. Months are delimited with thick black lines. Multiple years are tiled side by side.

Adherence and ridership are visualized in similar ways. For adherence, the color of a day is determined by the average of the absolute value of the difference between the scheduled arrival time and the actual arrival time. For ridership, the color for each day is determined by the average number of passengers who boarded on that day. Less extreme values on each scale (buses arriving close to their scheduled time and low number of passengers) are more blue, and more extreme values (buses arriving very late or early and high numbers of passengers) are more red. Days with missing data are left blank (white).

In the initial usability study, calendar view displayed years horizontally. Several users complained that this was unintuitive and at their suggestion, calendar view was redesigned vertically. A participant also noted that while the original color scheme (green to red) was arguably more intuitive, it it would be unacceptable for an individual who was color blind. Furthermore, the original color scheme had too many colors, which made it hard to distinguish one from another. We corrected this by changing the color scheme to use a spectrum safe for individuals who are colorblind with only five colors.

\subsubsection{The Trip Subview of Calendar View}
\begin{figure}[htb]
 \centering
 \includegraphics[width=8cm]{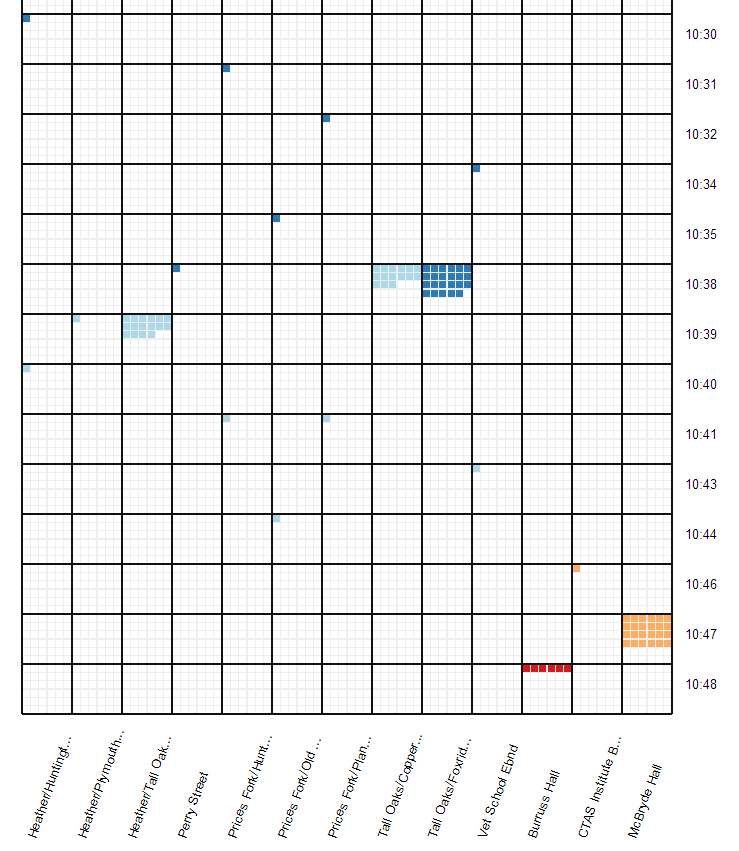}
 \caption{Trip subview of calendar view. The x axis contains each stop in a route, the y axis contains each trip made along that route for a particular day. Ridership is indicated by the number of colored squares with color corresponding to adherence.}
 \label{trip_component}
\end{figure}

The trip subview (Fig. \ref{trip_component}) of calendar view visualizes adherence and ridership for each trip of a route route that occurred on a specific day. Data is displayed on a rectangular grid with rows representing trips and columns representing stops. Each cell contains min(n, 36) squares where n is the number of passengers who boarded at that stop on that trip. \footnote{Displaying a maximum of 36 squares was chosen in order to display every boarding passenger most of the time (buses can only carry around 50 passengers at once) while allowing the squares to be large enough to easily count without taking up too much space.} The squares are colored according to adherence. Adherence is mapped to a spectrum from dark blue (bus arrived on time) to dark red (bus was very early or late). One of the most important goals of MetroViz was to be able to display the stops, routes, and trips where the most passengers were affected by a bus with poor adherence. Trip subview achieves this by attracting the user's eye to large red areas on the grid.

Originally, the trip subview displayed stops as circles with different sizes. Our first usability study revealed this to be unsuitable because it was too difficult to compare the sizes of circles. In addition, the original color scheme was inconsistent with the rest of the system and more difficult to use because it was a continuous spectrum. We addressed this problem by using the same color scheme as the calendar view and displaying the passengers as discrete quantities.

Unfortunately, even the improved trip subview of calendar view suffers from the quality of the Blacksburg bus system dataset. Many stops report no passengers boarding while others report unrealistic numbers of passengers boarding (sometimes as high as 160 - especially unlikely given that most buses can hold no more than 60 passengers).

\subsubsection{The Stop Subview of Calendar View}
\begin{figure}[htb]
 \centering
 \includegraphics[width=8cm]{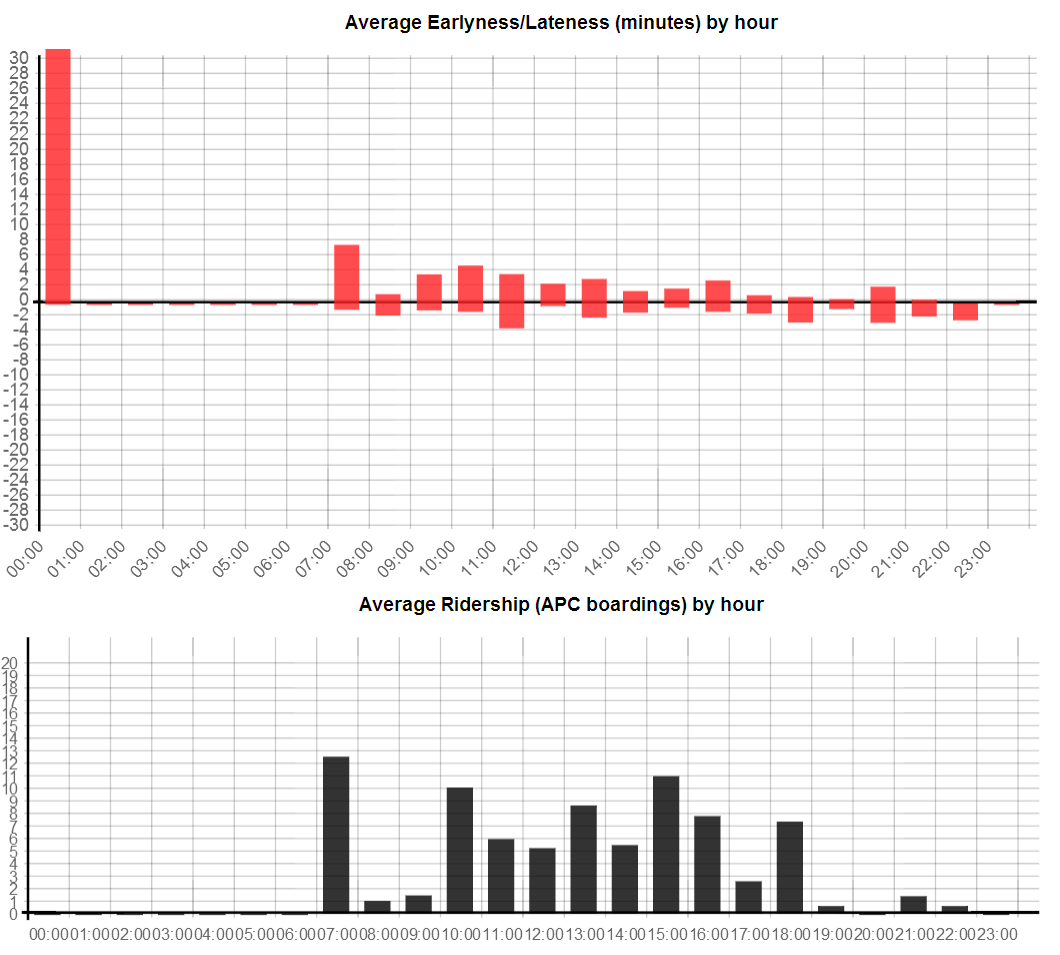}
 \caption{Stop subview of calendar view with average adherence in red and ridership in black. Note some bars with simultaneous positive and negative values - the negative value is the average earliness for all stops that entire hour (a late bus has earliness 0) and the positive value the average lateness. The total length of the resulting bar is equal to the average of the absolute value of the arrival delta.}
 \label{stop_component}
\end{figure}

The stop subview (Fig. \ref{stop_component}) of calendar view visualizes ridership and adherence of a single stop for a single day. One of the major challenges of the stop subview was displaying adherence data in a sensible and intuitive fashion at this level of detail. Unlike the other high-level views, we wanted to capture the distinction between earliness and lateness (e.g. a bus leaving 5 minutes earlier than expected may result in a significantly longer wait for the next bus; a bus leaving 5 minutes later than expected is only a 5 minutes wait.) This ruled out the strategy of taking the sum of the absolute value of the deltas between scheduled and actual arrival time as we did for calendar view and the trip subview. Averaging the deltas is also unsuitable - a bus that was 10 minutes early and and another bus that was 10 minutes late would average out to be exactly on time. In addition, we desired a display that would simultaneously show the number of passenger boardings to visualize the number of people affected by non-adherence.

Stop subview went through a number of iterations. The initial design showed the number of boardings and either earliness (in green) or lateness (in red) for every stop that day at a particular location. A variant of this design used the same color scheme, but binned stops by hour (the values shown for each hour are the average values for all stops that occurred during that hour) for a more predictable use of space and consistent spacing. Bucketing also had the effect of sometimes showing early (green) and late (red) bars for the same hour.

In the initial usability tests, we also tested versions of the above bar charts with adherence and ridership displayed on separate charts. In addition, there were line chart versions of all of the bar charts. Most of the initial usability test participants disliked the design choices in the initial bar charts and the line chart versions were universally disliked. The test participants also universally preferred bucketing the stops by hour instead of displaying every stop individually. The most common complaints were a dislike of different types of data (number of passengers and minutes early/late) displayed on the same scale and the display of both earliness and lateness as positive values on the chart.

We had several suggestions that earliness be displayed as a negative value and lateness be displayed as a positive one. Our first re-design kept the green and red bars separate and simply mirrored the earliness bar over the x-axis. Later, we realized that we could reduce space by stacking the early and late bars on top of each other and with a clear origin line, display the stacked bars as a single color. This display strategy also has the benefit that the total length of each bar is the same length of a bar displaying the sum of the absolute value of the deltas between scheduled and actual arrival time which helps maintains consistency with calendar and trip subview's adherence ranking system. A ridership chart is also shown directly below the adherence chart to help visualize the number of people affected by non-adherence.

Finally, we also investigated the possibility of using radar charts to display population or adherence levels as stop time naturally mapped to a clock-like display. However, we found that such a chart makes comparisons between hours difficult and in usability tests was generally rejected in favor of the simple bar chart.

\section{Evaluation}
\subsection{Methodology}
We decided that the best method for evaluating the performance of the MetroViz system was to conduct usability tests on several test subjects as opposed to a more formal controlled experiment - there were no concrete hypotheses to test, nor could we define suitable dependent and independent variables. Although domain experts were amongst our test subjects, the short duration of this project made a long term case study infeasible. Our usability tests involved a training session followed by a set of pre-determined tasks that the test subjects were asked to perform. We encourage participants to ``think out loud'' and give verbal feedback which we recorded during the test session. We also observed if the participant has any undue difficulty in performing the task. As mentioned previously, many of our design choices were based on early usability test suggestions and feedback.

Our usability test had four steps: First, we gave a demonstration of MetroViz individually. Second, we collected some information from the participant to determine if they had any prior domain knowledge on visualizing transportation data. Then, we gave the subject a set of tasks with seven different themes. We would assist the users verbally if they had extreme difficulty in performing a task. Finally, we administered a Likert Scale style questionnaire to obtain more formal feedback. 
\noindent
The tasks for the Usability Test were: 
\begin{enumerate}
  \item Given a bus-stop, what are the routes this stop is present in. What road in Blacksburg, Virginia is this stop located on? What are the most common passenger types (whether students, faculty or general public) that use this stop?
  \item Given the route view, what are the shortest and longest routes in terms of number of stops?
  \item Given two routes, what type of passengers mostly use the route?
  \item Given a single route, geographically, which part of Blacksburg, Virginia does this route serve?
  \item Given a bus stop, on what days was the arrivals at the stop unusually not on schedule? On what days were an unusually large number of passengers using the stop?
  \item Given a (bus stop, date) pair, what were the peak hours on the day? Was there any serious delay on the day, or was the bus too early at the stop?
  \item Given a (route, date) pair, what are the busiest stops on the route? At what time did bus service begin and retire for the day? What are the peak hours for the day, for that route?
\end{enumerate}

Tasks 1 to 4 test specifically, the map view and route view of MetroViz. We observed if the subject could use the search functionality correctly, if (s)he could scroll up or down the list of routes and navigate through the map geographically. We also tested if the user can make inferences on the types of passengers using the service (e.g. whether student or faculty/staff). Tasks 5 to 7 test if the subject is able to correctly interpret the stop and trip subviews of calendar view, and whether they are adept at reading the calendar. The user could make inferences like discovering particularly busy stops and routes or routes and stops that are under-utilized as well as discovering routes that are regularly not on schedule, and the number of passengers affected. Finally, after the usability tests, we administered a Likert Scale style questionnaire. and solicited feedback on the quality of each of the views (map, calendar, route, trip and stop). We also sought opinions on the color schemes we used, and whether there were any features that the users would like added to the system. We took note of any part of the system that the user found excessively difficult or frustrating. The list of tasks for usability testing and the feedback questionnaire and provided is available in the appendix. 

We gained some valuable insights upon performing two rounds of usability testing. The first round of usability testing was performed with three members of the CATT lab when MetroViz was at a more nascent stage. Since each of the test subjects had expertise in using visualization tools specifically for public transportation, they can be considered as domain experts. The second round of usability testing, unlike the previous round was performed with three domain experts and six non-experts.

\subsection{Results}

Based on the first round of usability testing with domain experts, the following changes were made to MetroViz :

\begin{enumerate}

\item \textbf{Map View and Route View:} The legend on the map view which showed fare types was made static (i.e, it would not move when the map view pans). We added a search functionality that would auto-complete and was case insensitive. The map view and route view used a similar color scheme for visual consistency. Tooltips were provided to highlight stop name across all stops. 

\item \textbf{Calendar View:} The orientation of the calendar view was changed from horizontal to vertical. The color scheme was changed from green-yellow-red to blue-white-red to make it more readable to color-blind users. We provided tool-tips that would display the date upon hovering in the calendar view.

\item \textbf{Stop Subview:} We decided to display two simple bar charts - one for adherence and the other for ridership. In the adherence bar chart, to differentiate better between buses that arrive early and buses that arrived late, we moved the x-axis to the center of the graph. A bar above the x-axis meant that buses arrived late and below the x-axis meant that buses arrived early on average at that stop. 
 
\item \textbf{Trip Subview:} We performed a drastic overhaul of the trip subview, incorporating a design based on squares to indicate ridership along with a blue-white-red color scheme to indicate adherence. The previous approach used circles of different radii, but this was scrapped because it was felt that comparing the radii of circles was difficult.

\begin{figure}[htb]

 \centering

 \includegraphics[width=8cm]{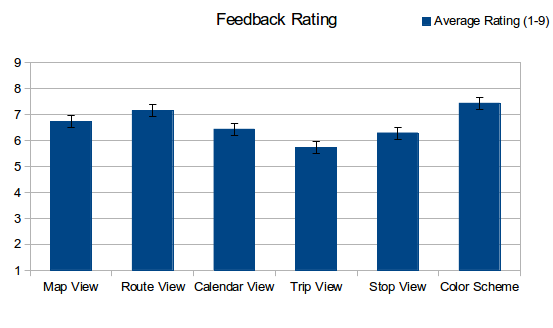}

 \caption{Results from the feedback questionnaire on the quality of map, route and calendar view, and trip and stop subview.}

 \label{ratings}

\end{figure}

\end{enumerate}

\noindent

The following changes were made to MetroViz on the basis of the second round of usability testing which involved both domain experts and non-experts. 

\begin{enumerate}

\item \textbf{Map View and Route View:} All our test subjects complained about a difficulty in hovering in the route view. Specifically, our system did not retain previously clicked stop names. We made the route view more user-friendly by intelligently enabling and disabling hover functions and retaining the most recently clicked stop name. 

\item \textbf{Stop Subview:} We added labels to the bar charts in the stop subview. 

\end{enumerate}

The results of the Likert Scale style questionnaire are presented in Fig. \ref{ratings}. We asked test participants to rate various features of our system on a scale of 1 to 9, where ``1" indicated a ``strong dislike", whereas ``9" indicated that the participants ``strong like" a feature. Based on the results, we can conclude that the participants found it easy to navigate through the stops in the map and route views, but found it somewhat difficult to interpret the visualization in the trip subview. Some of the recommendations made by the participants which can be implemented in the future are: 

\begin{enumerate}

\item[] \textbf{R1:} Make better use of screen real-estate by appropriately resizing the various views. Provide the users an option to disable one or more view. 

\item[] \textbf{R2:} Provide month and day-of-the-week labels in the calendar view which makes it easier for the participants to close in on a specific date. 

\item[] \textbf{R3:} The trip subview is extremely large, especially if a particular route has many trips per day, necessitating a lot of up and down scrolling. We could divide the trip subview into morning, afternoon, evening and night trips to reduce the size of the trip subview. A similar recommendation from another participant is to group by hour. 

\item[] \textbf{R4:} Eliminate the excessive use of scroll-bars. Despite our best efforts, we could not come up with a simple solution for this problem, even conceptually.

\item[] \textbf{R5:} Provide a control panel to enable sorting based on ridership and adherence.

\end{enumerate}
 
\section{Discussion}

Usability studies revealed that we have found a promising way to visualize adherence and ridership data. Participants were easily able to locate specific routes and stops, find unusual days, discover busy times of the day where the buses were not adhering to schedule, and complete other important tasks that a public transportation manager must perform. Despite our many successes, there is still much room for improvement. 

\subsection{Future Work}

\subsubsection{Data Quality}

Although three gigabytes is not exceptionally large, MetroViz suffers from many of the problems that plague ``big data'' - long load times and incomplete and inaccurate data. Calendar view handles missing data very well - a day without adherence or stop data is displayed as an unobtrusive blank square. Map view virtually ignores missing data - as along as a single fare count by fare type is available, the donut glyphs for each stop can be rendered. Unfortunately, trip subview does not handle missing data as well - during usage we often ran into problems where selecting a single day would bring up a very large number of trips without ridership data, resulting in a large amount of wasted space. Trip subview was mostly unsuccessful on the Blacksburg dataset, but given higher quality data, it may prove worthwhile in the future. It is impossible to definitively access the quality of the trip subview without higher quality data.

The long load times (up to 10 seconds for certain routes or stops) are somewhat unavoidable - calendar view requires the complete dataset for a stop/route over the entire time span available and preloading data would result in an excessively long startup time. The bottleneck appears to be the data processing done in javascript - in the future, delegating the time intensive computations to a faster backend language may help mitigate this problem.

\subsubsection{Visualization}

Managing limited screen real estate has been very difficult given the large number of views we provide the user. Although we had some successes according to the participants in the usability study (most notably pinning the calendar view to the right side of the screen), the lack of sufficient screen space was dealt with less gracefully elsewhere with excessive use of scroll bars. The route view and trip subview both extend too far when displaying long routes. This could potentially be resolved by splitting long routes onto multiple lines. Despite stacking the early and late bars, the stop subview charts required a large amount of space to get precise readings on adherence and ridership levels - this made simultaneously comparing multiple days impractical on standard screen sizes. Either larger screen sizes or dynamically resizable charts would be needed to mitigate this shortcoming. However, the ability to visualize earliness, lateness and total adherence in a single chart may be worth the cost of extra screen space.

In addition, there are still opportunities to leverage unused attributes of the dataset. MetroViz does not currently incorporate data about bus drivers into its visualization. In future versions, it would be very useful it was possible to filter the data by driver. Another potential avenue would be to introduce separators in the trip subview to show where one bus driver stopped and another started.

Finally, the most important feature to be introduced to future versions of MetroViz is the ability to sort routes and stops by adherence and ridership. Although MetroViz is very effective at locating temporal regions of poor adherence and high ridership, it is not as effective at identifying routes and stops with exceptional values for both adherence and ridership. Allowing the user to sort routes and stops would be a major step towards addressing this problem.

\section{Credits}
\begin{itemize}
  \item \textbf{Fan:} Coding for map view; wrote sections: abstract, introduction, related works, data modeling; edited video; coordinated with CATT Lab.
  \item \textbf{Josh:} Coding for stop subview; wrote/edited visual design sections; proofreading; narrated/recorded video.
  \item \textbf{Peter:} Coding for calendar view, trip subview, and interfaces between several views; various writing and editing;
  \item \textbf{Varun:} Coding for route view; aggregating usability study data; writing usability study sections
  \item \textbf{Yoav:} Data preprocessing; SQL server; webserver coding.
\end{itemize}

\section{Acknowledgments}
The authors would like to thank Michael VanDaniker (CATT lab) for providing our dataset and his invaluable advice regarding design choices. We would also like to thank members of the CATT Lab, and our friends for volunteering in the usability tests and providing invaluable feedback. 

\bibliographystyle{abbrv}
\bibliography{template}
\end{document}